\documentclass[acmsmall]{acmart}
\acmConference[arxiv]{arxiv pre-print versions}{2024}{}
\renewcommand\footnotetextcopyrightpermission[1]{}

\usepackage{amsmath, amsfonts}
\usepackage{algorithmic}
\usepackage{graphicx}
\usepackage{textcomp}
\usepackage{xcolor}
\usepackage{listings,listings-rust}
\usepackage{caption,subcaption}
\usepackage{enumerate}
\usepackage{multirow}
\usepackage{color}
\usepackage{colortbl}
\usepackage[ruled,vlined]{algorithm2e}
\usepackage{amsfonts}
\usepackage{threeparttable}
\usepackage{booktabs}
\usepackage{makecell}
\usepackage{cleveref}
\usepackage{soul}

\definecolor{gray_blank}{gray}{.7}
\definecolor{gray_item}{gray}{.7}

\definecolor{generic}{rgb}{0.75, 0, 0}
\definecolor{mut}{rgb}{0, 0.5, 0}
\definecolor{primitive}{rgb}{0, 0, 0.5}
\definecolor{oldspn}{rgb}{0, 0, 0.75}
\definecolor{oldspb}{rgb}{30, 144, 255}

\newtheorem{goal}{\textbf{GOAL}}[section]
\newtheorem{filter}{\textbf{FILTER}}[section]

\crefname{goal}{Goal}{Goal}
    
\begin{document}

\title{Fearless Unsafe. A More User-friendly Document for Unsafe Rust Programming Base on Refined Safety Properties.}

\author{Mohan Cui}
\affiliation{%
  \institution{School of Computer Science, Fudan University}
  \country{China}
}
\email{cuimohan@fudan.edu.cn}

\author{Penglei Mao}
\affiliation{%
  \institution{School of Computer Science, Fudan University}
  \country{China}
}
\email{plmao22@m.fudan.edu.cn}

\author{Shuran Sun}
\affiliation{%
  \institution{School of Computer Science, Fudan University}
  \country{China}
}
\email{srsun20@fudan.edu.cn}

\author{Yangfan Zhou}
\affiliation{%
  \institution{School of Computer Science, Fudan University}
  \country{China}
}
\email{zyf@fudan.edu.cn}

\author{Hui Xu}
\authornote{Corresponding author.}
\affiliation{%
  \institution{School of Computer Science, Fudan University}
  \country{China}
}
\email{xuh@fudan.edu.cn}

\begin{abstract}
Rust, a popular systems-level programming language, has garnered widespread attention due to its features of achieving run-time efficiency and memory safety. With an increasing number of real-world projects adopting Rust, understanding how to assist programmers in correctly writing unsafe code poses a significant challenge. Based on our observations, the current standard library has many unsafe APIs, but their descriptions are not uniform, complete, and intuitive, especially in describing safety requirements. Therefore, we advocate establishing a systematic category of safety requirements for revising those documents. 

In this paper, we extended and refined our study in ICSE 2024. We defined a category of \textit{Safety Properties} (22 items in total) that learned from the documents of unsafe APIs in the standard library. Then, we labeled all public unsafe APIs (438 in total) and analyzed their correlations. Based on the safety properties, we reorganized all the unsafe documents in the standard library and designed a consultation plugin into rust-analyzer as a complementary tool to assist Rust developers in writing unsafe code. To validate the practical significance, we categorized the root causes of all Rust CVEs up to 2024-01-31 (419 in total) into safety properties and further counted the real-world usage of unsafe APIs in the crates.io ecosystem. 
\end{abstract}


\maketitle

\section{Introduction}
Rust is a popular system programming language designed for achieving memory safety and efficiency~\cite{matsakis2014rust, klabnik2019rust}. This is attributed to extensive static analysis during compilation~\cite{fulton2021benefits}; various zero-cost abstractions collectively provide those guarantees~\cite{zhu2022learning}, and the typical examples are ownership and generics~\cite{jung2017rustbelt}. Limited by using static analysis in the compiler, programmers must invest significant effort in writing code that can satisfy the syntax requirement to pass the compilation, leading to a steep learning curve~\cite{fulton2021benefits}. However, due to the enticing rewards~\cite{popular}, Rust has attracted numerous real-world projects to code translations, especially for the system programs~\cite{shen2020occlum,levy2017tock,levy2017multiprogramming}. Since 2016, Rust has consistently been the most popular programming language in the open-source community~\cite{survey16,survey17,survey18,survey19,survey20,survey21,survey22}.

\begin{figure}[]
\begin{subfigure}[]{\textwidth}
\begin{lstlisting}[language=Rust, style=colouredRust, label=list:take, caption=Source code of ManuallyDrop::take in Rust std.\\]
// impl<T> ManuallyDrop<T>
pub unsafe fn take(slot: &mut ManuallyDrop<T>) -> T {
    // SAFETY: we are reading from a reference, which is
    // guaranteed to be valid for reads.
    unsafe { ptr::read(&slot.value) }
}
\end{lstlisting}
\label{fig:relation1}
\end{subfigure}
\begin{subfigure}[]{\textwidth}
\begin{lstlisting}[style=docs, label=list:takedoc, caption=Document of ManuallyDrop::take in Rust 1.75.\\]
Takes the value from the ManuallyDrop<T> container out.
This method is primarily intended for moving out values in drop. Instead of using ManuallyDrop::drop to manually drop the value, you can use this method to take the value and use it however desired.
Whenever possible, it is preferable to use into_inner instead, which prevents duplicating the content of the ManuallyDrop<T>.
Safety
This function semantically moves out the contained value without preventing further usage, leaving the state of this container unchanged. It is your responsibility to ensure that this ManuallyDrop is not used again.
\end{lstlisting}
\end{subfigure}
\caption{Example of an unsafe API in Rust std. Listing~\ref{list:take} provides the source code of an unsafe method of struct ManuallyDrop<T>. Listing~\ref{list:takedoc} extracts the document in Rust 1.75. This document introduces the usage, functionality, and safety requirements to comply with by using a section labeled \textit{Safety}.}
\label{fig:take}
\end{figure}

\begin{table*}[]
\caption{A list of APIs with the same side effects and their document slices in Rust 1.75. These APIs input mutable pointers and return a typed owner. In previous research, it has been reported that misuse of these APIs can incur double free. The \textbf{\textit{Safety}} section is highlighted, the description of the safety requirement is underlined, and the side effect is bolded.}
\label{table:unsafe_ctors}
\resizebox{\linewidth}{!}{
\begin{tabular}{c|c|p{10cm}c}
    \toprule[1pt]
	\textbf{Implemented Type}    & \textbf{Unsafe Method} & \textbf{Safety Description Slices in API Documents.} \\
	\midrule[1pt]
	\multirow{4}{*}{\textbf{impl}<\textcolor{generic}{\textbf{\textcolor{generic}{\textbf{T: ?Sized}}}}> \textcolor{mut}{*mut} \textcolor{generic}{\textbf{T}}} & \multirow{4}{*}{\textbf{fn} \textbf{read}(\textbf{self}) -> \textcolor{generic}{\textbf{T}}} & read creates a bitwise copy of T, regardless of whether T is Copy. \ul{If T is not Copy, using both the returned value and the value at *src can \textbf{violate memory safety}.} Note that assigning to *src counts as a use because it will attempt to drop the value at *src.\\
	\hline
	\multirow{4}{*}{\textbf{impl}<\textcolor{generic}{\textbf{T}}> \textbf{ManuallyDrop}<\textcolor{generic}{\textbf{T}}>} & \multirow{4}{*}{\textbf{fn} \textbf{take}(\textcolor{mut}{\&mut} \textbf{ManuallyDrop}<\textcolor{generic}{\textbf{T}}>) -> \textcolor{generic}{\textbf{T}}} & \textbf{\textit{Safety}}: This function semantically moves out the contained value without preventing further usage, leaving the state of this container unchanged. It is your responsibility to \ul{ensure that this ManuallyDrop is not used again.} \\
	\hline
	\multirow{3}{*}{\textbf{impl}<\textcolor{generic}{\textbf{T: ?Sized}}> \textbf{Box}<\textcolor{generic}{\textbf{T}}>} & \multirow{3}{*}{\textbf{fn} \textbf{from\_raw}(\textcolor{mut}{*mut} \textcolor{generic}{\textbf{T}}) -> \textbf{Self}} & This function is unsafe because improper use may lead to \textbf{memory problems}. For example, \ul{a \textbf{double-free} may occur if the function is called twice on the same raw pointer.}\\
	\hline
	\multirow{4}{*}{\textbf{impl}<\textcolor{generic}{\textbf{\textcolor{generic}{\textbf{T: ?Sized}}}}> \textbf{Rc}<\textcolor{generic}{\textbf{T}}>} & \multirow{4}{*}{\textbf{fn} \textbf{from\_raw}(\textcolor{mut}{*const} \textcolor{generic}{\textbf{T}}) -> \textbf{Self}} & \ul{The raw pointer must have been previously returned by a call to Rc<U>::into\_raw where U must have the same size and alignment as T.} The user of from\_raw has to \ul{make sure a specific value of T is only dropped once.} \\
	\hline
	\multirow{4}{*}{\textbf{impl}<\textcolor{generic}{\textbf{\textcolor{generic}{\textbf{T: ?Sized}}}}> \textbf{Arc}<\textcolor{generic}{\textbf{T}}>} & \multirow{4}{*}{\textbf{fn} \textbf{from\_raw}(\textcolor{mut}{*const} \textcolor{generic}{\textbf{T}}) -> \textbf{Self}} & \ul{The raw pointer must have been previously returned by a call to Arc<U>::into\_raw where U must have the same size and alignment as T.} The user of from\_raw has to \ul{make sure a specific value of T is only dropped once.} \\
	\hline
	\multirow{4}{*}{\textbf{impl} \textbf{CString}} & \multirow{4}{*}{\textbf{fn} \textbf{from\_raw}(\textcolor{mut}{*mut} \textcolor{primitive}{\textbf{c\_char}}) -> \textbf{Self}}  & \textbf{\textit{Safety}}: This should only ever be called with \ul{a pointer that was earlier obtained by calling CString::into\_raw.} Other usage (e.g., \ul{trying to take ownership of a string that was allocated by foreign code}) is likely to lead to \textbf{undefined behavior} or \textbf{allocator corruption}.\\
	\hline
	\multirow{4}{*}{\textbf{impl}<\textcolor{generic}{\textbf{T}}> \textbf{Vec}<\textcolor{generic}{\textbf{T}}>} & \multirow{4}{*}{\textbf{fn} \textbf{from\_raw\_parts}(\textcolor{mut}{*mut} \textcolor{generic}{\textbf{T}}, \textcolor{primitive}{\textbf{usize}}, \textcolor{primitive}{\textbf{usize}}) -> \textbf{Self}} & \textbf{\textit{Safety}}: The ownership of ptr is effectively transferred to the Vec<T> which may then deallocate, reallocate or change the contents of memory pointed to by the pointer at will. \ul{Ensure that nothing else uses the pointer after calling this function.}\\
	\hline
    \multirow{4}{*}{\textbf{impl} \textbf{String}} & \multirow{4}{*}{\textbf{fn} \textbf{from\_raw\_parts}(\textcolor{mut}{*mut} \textcolor{primitive}{\textbf{u8}}, \textcolor{primitive}{\textbf{usize}}, \textcolor{primitive}{\textbf{usize}}) -> \textbf{Self}} & \textbf{\textit{Safety}}: The ownership of buf is effectively transferred to the String which may then deallocate, reallocate or change the contents of memory pointed to by the pointer at will. \ul{Ensure that nothing else uses the pointer after calling this function.}\\
    \bottomrule[1pt]
\end{tabular}
}
\end{table*}

As more programs employ Rust, correctly writing unsafe code becomes crucial for the safety promise~\cite{astrauskas2020programmers}. Safety isolation is one of the innovations introduced by Rust~\cite{qin2020understanding}. It segregates parts where the compiler can ensure memory safety into Safe Rust and uses \textbf{\textit{unsafe}} keyword as a superset that is not fully constrained by the compile-time check~\cite{klabnik2019rust,unsafesyntactic}. Under this isolation, the code containing unsafe operations must be encapsulated within an unsafe block; otherwise, it will fail to compile. Due to the absence of strict compiler checks within the unsafe scope, developers may perceive unsafe code as a backdoor to resolve the compiler error, which is error-prone to incurring undefined behavior (UB).

In this paper, we extended and refined our study in ICSE 2024. To study this topic, the standard library, which has a diverse collection of unsafe APIs and comprehensive documents, stands out as an ideal target~\cite{ruststd}. Each unsafe API in std-lib has one section named \textbf{\textit{Safety}} that outlines its safety requirements. As shown in Figure~\ref{fig:take}, \texttt{ManuallyDrop::take}~\cite{take} specifies that users cannot reuse the container after calling this function. Table~\ref{table:unsafe_ctors} presents our observations that the safety requirements are neither uniform nor complete. Recent studies revealed misusing a set of unsafe APIs that would cause double free~\cite{xu2021memory, cui2023safedrop, bae2021rudra}, as listed in Table~\ref{table:unsafe_ctors}. Despite having the same safety requirements and side effects, the documents do not exhibit significant uniformity. Also, the descriptions are not complete, only \texttt{Box<T>} explicitly mentioning the risk for double free.

Another observation is the lack of intuitiveness. Utilizing the \textbf{\textit{Safety}} section to enumerate safety requirements aligns with intuition. The majority of APIs listed adheres to this standard, such as \texttt{String}, \texttt{Vec<T>}, \texttt{CString}, and \texttt{Box<T>} in Table~\ref{table:unsafe_ctors}. However, \texttt{Rc<T>} and \texttt{Arc<T>} miss the \textbf{\textit{Safety}} section, and issues arising from \texttt{ptr::read}~\cite{read} are stated in an external section: \textit{Ownership of the Returned Value}. Simultaneously, multi-level jumps require users to click hyperlinks frequently. As seen in Listing~\ref{list:link1}, readers of the documentation for \texttt{impl *const T}: \texttt{read} need to navigate to \texttt{ptr::read}; in Listing~\ref{list:link2}, the definition of \textit{\textbf{valid pointer}} is documented in another lengthy page, demanding users to invest significant effort in reading them.

We advocate for the Rust community to develop a unified, complete, intuitive, and well-categorized document system for unsafe APIs to help users write unsafe code correctly. Specifically, this paper aims to address the following research questions (RQs):

\begin{enumerate}[0]
\item[$\bullet$] \noindent \textbf{RQ-1.} How to summarize finer-grained safety properties (requirements) that should be satisfied in Unsafe Rust? (\S\ref{sec:sp})
\item[$\bullet$] \noindent \textbf{RQ-2.} If the safety properties can be used to provide user-friendly documents for Unsafe Rust programmers? (\S\ref{sec:plugin})
\item[$\bullet$] \noindent \textbf{RQ-3.} Can those safety properties cover the root causes of the existing program vulnerabilities caused by Unsafe Rust? (\S\ref{sec:cveio})
\end{enumerate}

\begin{figure}[]
\begin{subfigure}[]{\textwidth}
\begin{lstlisting}[style=docs, label=list:link1, caption=Document of *const T::read in Rust 1.75.\\]
Primitive Type pointer
pub const unsafe fn read(self) -> T

Reads the value from self without moving it. This leaves the memory in self unchanged.
See ptr::read for safety concerns and examples.
\end{lstlisting}
\end{subfigure}
\begin{subfigure}[]{\textwidth}
\begin{lstlisting}[style=docs, label=list:link2, caption=Document of ptr::read in Rust 1.75.\\]
std::ptr::read
pub const unsafe fn read<T>(src: *const T) -> T

Reads the value from src without moving it. This leaves the memory in src unchanged.

Safety
Behavior is undefined if any of the following conditions are violated:
    src must be valid for reads.
    src must be properly aligned. Use read_unaligned if this is not the case.
    src must point to a properly initialized value of type T.
Note that even if T has size 0, the pointer must be non-null and properly aligned.
\end{lstlisting}
\end{subfigure}
\caption{Two instances of non-intuitiveness in the std unsafe documents. These two functions are named \textit{read}, with identical functionality and signature. The document in List~\ref{list:link1} requires a jump to List~\ref{list:link2}. However, List~\ref{list:link2} contains a hyperlink leading to a secondary jump toward the informal definition of \textit{valid pointer}.}
\end{figure}

For each RQ, we designed different experiments centered around unsafe documents. This paper is an extended version of our conference paper \textit{"Is unsafe an Achilles’ Heel? A Comprehensive Study of Safety Requirements in Unsafe Rust Programming"}~\cite{cui2024unsafe}. To answer RQ1, we reviewed all public unsafe APIs from the standard library~\cite{ruststd}. We refined our safety properties defined in the previous work, labeled the involved APIs, and conducted correlation analysis. To answer RQ2, we revised the unsafe documents in the standard library, grouping the document slice, safety property, and parameter into triplets. We integrated the revised documents into rust-analyzer as a complementary plugin for users. To answer RQ3, we examined all Rust CVEs up to 2024-01-31 from CVEmitre~\cite{cve}, filtered them based on the root causes, and classified them according to the violated safety properties. Subsequently, we analyzed the distribution of unsafe APIs within the crates.io~\cite{cratesio}. 

We finally reviewed 438 unsafe APIs and defined 13 safety properties with 22 sub-properties, categorizing them into preconditions or postconditions. We conducted the correlation analysis on the labeled APIs and the result demonstrated the safety properties for dereferencing. Next, we reorganized all the documents by analyzing the parameters corresponding to safety properties and document slices. The revised documents have been incorporated into rust-analyzer, and it can assist users by displaying a detailed list of the triplets while hovering over the source code. Finally, we classified current CVEs based on safety properties, with 198 out of 419 originating from misusing unsafe code. Notably, 86.36\% of these errors were attributed to misuse of std APIs. Therefore, we conducted a statistical analysis of std unsafe API usage within the Rust ecosystem, comprising 103,516 libraries on crates.io. 

Our main contributions are listed as follows:
\begin{enumerate}[0]
\item[$\bullet$] \noindent We conducted the empirical study by analyzed unsafe API documents from the standard library to establish a systematic categorization of safety requirements for Unsafe Rust.
\item[$\bullet$] \noindent We summarized 13 safety properties with 22 sub-properties and labeled all unsafe APIs through this criteria, and the labeled data were evaluated via correlation analysis.
\item[$\bullet$] \noindent We reorganized all documents of the unsafe functions in the standard library and integrated them into the rust-analyzer, making it open source.
\item[$\bullet$] \noindent We categorized all Rust CVEs based on safety properties, forming a collection that can serve as a benchmark, and we collected API usage statistics within crates.io to understand the frequency.
\end{enumerate}

\section{Background}

\subsection{Safety Isolation in Rust}
Unsafe Rust is a superset of Safe Rust~\cite{qin2020understanding}. Safe Rust employed static analysis to guarantee memory safety. Thus, undefined behavior cannot be triggered solely using Safe Rust~\cite{astrauskas2020programmers}. Consequently, it cannot be a system programming language due to the lack of low-level controls (\textit{e.g.,} manual memory management). Unsafe Rust is an essential design to compensate for this weakness~\cite{rustonomicon}. It is used if users need to interact with OS, hardware, or other programming languages to achieve better performance. A recent empirical study investigated the learning challenges of safety mechanisms via Stack Overflow and user surveys~\cite{zhu2022learning}. It was reported that Rust's safety mechanisms could be more learner-friendly but are restricted to Safe Rust. Instead, learning and utilizing Unsafe Rust may be a prerequisite for advanced Rust developers, especially for system programming. 

\subsubsection{\textit{unsafe} Keyword} \textit{unsafe} keyword is allowed to be used as a prefix for different contexts. The first scenario marks definitions with extra safety conditions (\textit{e.g.,} \textit{unsafe fn}, \textit{unsafe trait}). It indicates safety requirements the callers or implementations must uphold, which the compiler does not check. The second scenario marks code that needs to satisfy extra safety conditions (\textit{e.g.,} \textit{unsafe \{ \}}, \textit{unsafe impl}.). It signifies the requirements that the programmers should be seriously examined to satisfy the safety conditions inside the block. \textit{unsafe} keyword states that all unsafe operations must be encapsulated within unsafe scope. Notably, the trust between safe and unsafe parts is asymmetric~\cite{rustonomicon}. Manual inspection is required to ensure the safe portion's data adheres to the unsafe operations' contracts. Conversely, the safe code should keep the unconditional belief that the unsafe part is always correct.

\subsubsection{\textit{unsafe} Operations} Safety isolation is designed for different aims. Safe Rust is a safe programming language for tasks that do not require low-level interactions. Contrariwise, Unsafe Rust fully leverages the capabilities of a system programming language. Unlike C/C++, Unsafe Rust still requires adherence to certain contracts from the safe portion, such as ownership. The main differences are five unsafe operations are allowed to be used only in unsafe blocks: (1) Dereference raw pointers; (2) Call unsafe functions; (3) Implement unsafe traits; (4) Mutate static variables; and (5) Access fields of unions~\cite{rustonomicon}. These operations provide flexibility but come with the risk of incurring undefined behavior and the responsibility of the users to review the correctness.

\subsection{Undefined Behavior in Rust}\label{sec:ub}
The undefined behavior in Rust is listed below. It is limited because this categorization targets the side effects of unsafe operations. Since no formal model of Rust's semantics defines precisely what is and is not permitted in unsafe code~\cite{reference}, additional behavior may be deemed vulnerable. In this paper, we add the underlined issues as program vulnerabilities if they are triggered by unsafe code: 

\begin{enumerate}[0]
\item[$\bullet$] \noindent Dereference (using the \texttt{*} operator on) dangling or unaligned raw pointers.
\item[$\bullet$] \noindent Break the pointer aliasing rules. References and boxes must not be dangling while they are alive.
\item[$\bullet$] \noindent Call a function with the wrong call ABI or unwinding from a function with the wrong unwind ABI.
\item[$\bullet$] \noindent Execute code compiled with platform features that the current thread of execution does not support.
\item[$\bullet$] \noindent Produce invalid values, as explained in Table~\ref{table:ub}, even in private fields and locals.
\item[$\bullet$] \noindent Mutate immutable data. All data inside a const item, reached through a shared reference or owned by an immutable binding, is immutable.
\item[$\bullet$] \noindent Cause data races.
\item[$\bullet$] \noindent Invoke undefined behavior via compiler intrinsics.
\item[$\bullet$] \noindent Use inline assembly incorrect.
\item[$\bullet$] \noindent \underline{Cause a memory leak and exiting without calling destructors.}
\item[$\bullet$] \noindent \underline{Trigger an unreachable path then aborting (or panicking).}
\item[$\bullet$] \noindent \underline{Arithmetic overflow.}
\end{enumerate}

\begin{table}[]
\caption{Invalid value for Rust types, alone or as a field of a compound type.}
\label{table:ub}
\resizebox{\linewidth}{!}{
\begin{tabular}{cp{13cm}c}
    \toprule[1pt]
	\textbf{Rust Type} & \textbf{Invalid Value}\\
    \midrule[1pt]
    \textbf{!} & Invalid for all values.\\
	\hline
	\textbf{\texttt{bool}} & Not \texttt{0} or \texttt{1} in bytes.\\
	\hline
	\textbf{\texttt{char}} & Outside \texttt{[0x0, 0xD7FF]} \& \texttt{[0xE000, 0x10FFFF]}.\\
	\hline
	\textbf{\texttt{str}} & Has uninitialized memory.\\
	\hline
	\textbf{numeric \texttt{i*/u*/f*}} & Reads from uninitialized memory.\\
	\hline
	\textbf{\texttt{enum}} & Has an invalid discriminant.\\
	\hline
	\textbf{reference} & Dangling, unaligned, or pointing to an invalid value.\\
	\hline
	\textbf{raw pointer} & Reads from uninitialized memory.\\
	\hline
	\textbf{\texttt{Box}} & Dangling, unaligned, or pointing to an invalid value.\\
	\hline
	\textbf{\texttt{fn} pointer} & NULL. \\
	\hline
	\multirow{3}{*}{\textbf{wide reference}} & Has invalid metadata. \texttt{dyn} \texttt{Trait} is invalid if it is not a pointer to a vtable for \texttt{Trait} that matches the actual dynamic trait the pointer or reference points to, and \texttt{slice} is invalid if the length is not a valid \texttt{usize}. \\
	\hline
	\textbf{custom type} & Has one of those custom invalid values. \\
    \bottomrule[1pt]
\end{tabular}
}
\end{table}

These undefined behaviors serve as the basis for classifying safety requirements. Other errors fall outside the scope (\textit{e.g.,} deadlocks and logic errors).

\section{Reviewing Rust Documents}\label{sec:sp}
To answer RQ1, this section defines a systematic procedure of extracting the categorization of safety requirements named \textbf{Safety Properties}, which is learned from the documentation of unsafe API in the standard library~\cite{ruststd}.

\subsection{Where to Obtain Safety Requirements?}
\texttt{Rustdoc}~\cite{rustdoc} is the official document system embedded in Rust programs. Programmers can use it to document the functionalities, calling requirements, side effects, and sample snippets in different scales. As for unsafe APIs, the safety requirements of arguments and return values must be explicitly specified in the document. Considering the quality of the documents in the ecosystem, we utilized the unsafe APIs within the standard library as our learning sample.

\subsubsection{Design Goals} Table~\ref{table:unsafe_ctors} reveals the lack of uniformity, completeness, and intuitiveness even in the standard library. Thus, we set the following goals to establish a systematic categorization of finer-grained safety requirements for Rust documents. The safety properties aim to provide a comprehensive and practical understanding of the safety considerations associated with Unsafe Rust.

\begin{goal}\label{g:gen}
	\textbf{Generality}: SP is not specific to one particular API's intricacies.
\end{goal}

\begin{goal}\label{g:amb}
	\textbf{Unambiguous}: SP adopts Rust's existing terminology and explanations.
\end{goal}

\begin{goal}\label{g:olp}
	\textbf{Nonoverlapping}: SP does not overlap, although they may be correlated.
\end{goal}

\begin{goal}\label{g:com}
	\textbf{Composability}: An Unsafe API's safety requirements can comprise several SPs.
\end{goal}

\begin{goal}\label{g:ess}
	\textbf{Essentiality}: Failure to comply with any SP would cause undefined behavior.
\end{goal}

\begin{goal}\label{g:prac}
	\textbf{Practicality}: SP must be seriously considered in real-world programming scenarios.
\end{goal}

\begin{goal}
	\textbf{Unilingual}: SP disregards FFI and the requirements of other programming languages.
\end{goal}

\subsubsection{Pre-filtering} The Rust standard library is a wide concept; it has several namespaces, including \texttt{core}, \texttt{std}, and \texttt{alloc}. The functions have redundancy in those namespaces, \textit{e.g.,} \texttt{std} and \texttt{core} are overlapping. Thus, we performed a pre-filtering for all unsafe APIs within \texttt{std}/\texttt{core}/\texttt{alloc} in Rust 1.75, considering both stable and nightly channels:

\begin{filter}
	For the methods exposed by both \texttt{core}, \texttt{std} and \texttt{alloc}, we kept only one of them.
\end{filter}

\begin{filter}\label{filter:num}
	For methods belonging to similar numeric types, we kept only one implementation.
\end{filter}

\begin{filter}
	For compiler intrinsics, we retained only those with no stable counterpart.
\end{filter}

We finally obtained an unsafe API collection comprising 438 unsafe functions, with 127 being folded as 11 unique APIs by Filter~\ref{filter:num} (\textit{e.g.,} \texttt{unchecked\_mul::<u8>/<u16>}~\cite{mulu8,mulu16} are merged). We streamlined an unsafe API queue for further analysis by applying the pre-filtering step.

\subsection{How to Extract Safety Properties?}
We performed a code review for all APIs in this collection to summarize what safety properties correspond with our design goals. Five rounds of code review finally constructed a set of safety properties with two categories and 22 items; the whole collection is shown in Table~\ref{table:sp}. For each API, we audited five sections: functional descriptions, safety requirements, subchapters, example snippets, and source code with comments.

\subsubsection{First Round} \underline{\textit{R1: SP Construction.}} We constructed a safety property category for API labeling in the first round. All the authors participated in the construction procedure. We adopted a sequential process by using a queue of API that requires auditing. For 30 APIs as a batch, the first author extracts potentially unrevealed safety properties, and the other authors verify them, which forms an assembly line. The first author cannot start a new batch unless the verification is completed.

\noindent \underline{\textit{SP Creation.}}
The category started as an empty set. The first author creates safety properties that others can modify. A new item will be defined whenever a safety description cannot be categorized under the existing scheme. Whenever the author suggests creating a new item, the decision is reached through a joint review. The names of safety properties are summarized from the document slices; only \textit{DualOwned} is the novel concept introduced in this paper.

\noindent \underline{\textit{SP Modification.}}
The author can submit a pull request to modify any safety properties (including the definition refinement, merging, splitting, or deleting items), and the approval is contingent on the joint review. For example, \textit{OverflowCheck} and \textit{IndexCheck} have been merged into \textit{Bounded}. Another example is splitting \textit{DualOwned} from \textit{Aliased} because they focused on objects and pointers, respectively.

\subsubsection{Second \& Third Round.}
After categorizing safety properties, the second and third rounds are dedicated to labeling the corresponding safety properties for each unsafe API. Since our labeling method employs a referenced table, which includes the definition and the document examples as the ground truth, conflicts during the labeling process are few.

\noindent \underline{\textit{R2: API Labeling.}} The second round created the initial labels for unsafe APIs. The first and second authors audited concurrently for all unsafe APIs in this round. Both authors conducted parallel reviews until all APIs were completed, then resolved conflicts in safety property labels.

\noindent \underline{\textit{R3: Labeling Review.}}
The process and conflict resolution in the third round of the audit are completely the same as in the second round. It serves solely for the secondary confirmation, correction, and supplementation of safety property labels.

\subsubsection{Fourth \& Fifth Round.} 19 safety properties have been defined until R3 is completed. Our previous work conducted a user survey to determine safety properties regarding precision, significance, usability, and frequency~\cite{cui2024unsafe}. As for precision, we found that safety properties defined with longer narratives generally had lower scores, especially for \textit{Consistent Layout} and \textit{Aliasing \& Mutating}. In this paper, we revised the safety properties, refining their contents.

\noindent \underline{\textit{R4: SP Revision.}} The SP revision involved changes to the whole categorization and labels, following the guidelines in R1-R3. This resulted in 13 safety properties comprising 22 sub-classes, with more detailed reasons for the change described in Section~\ref{sec:changes}.

\noindent \underline{\textit{R5: Document Revision.}} We revised the documents based on safety properties after getting the final categorization. The revised document connected the document slices with safety properties and corresponding arguments in the signature, forming a more formal version.

\noindent \underline{\textit{Conflict Resolution.}}
In cases of conflicts in SP creation (R1), SP modification (R1, R4), and SP labeling (R2, R3, R4), a joint review is initiated. It involves all the authors voting together; the action is implemented only if all the authors vote in agreement for SP creation or SP modification and more than three authors vote in agreement for SP labeling. Otherwise, the original decision is retained.

\section{Classifying Safety Requirements}\label{sec:plugin}
To answer RQ2, this section gives the classification of safety properties. We also elaborate on the refinement of the classification defined in our previous paper~\cite{cui2024unsafe}. Then we present how we reorganize the unsafe documents, building upon the safety properties. Finally, we introduce our plugin's framework that can help write unsafe code.

\subsection{What Safety Properties Should We Satisfy?}
We divided the safety properties into two categories based on the state of the function execution as in formal methods~\cite{berdine2005symbolic}, with no overlap between the sub-properties.

\begin{table*}[]
\caption{Safety properties learned from unsafe API documents in Rust standard library. Pre-condition and post-condition are two primary categories, and they have 13 items in total. We present the sum of the labeled unsafe API for each safety property and provide a detailed definition. Each safety requirement was also given a typical unsafe API as an example. Note that each SP may contain various sub-classes. The symbols \(\textcolor{red}{\blacktriangle}\) and \(\textcolor{blue}{\blacktriangledown}\) represent the satisfaction of hierarchy between the superclass and subclass.}
\label{table:sp}
\resizebox{\linewidth}{!}{
\begin{threeparttable}
\begin{tabular}{|c|c|p{11cm}|c|c}
    \toprule[1pt]
	\textbf{Safety Property} & \textbf{\#} & \textbf{Definition and the safety requirement of each Safety Property.} & \textbf{Unsafe API Example} \\	
	\hline
	\rowcolor{gray_blank} \multicolumn{4}{|c|}{}\\
	\rowcolor{gray_blank} \multicolumn{4}{|c|}{\multirow{-2}{*}{\textbf{Precondition Safety Property}}}\\
	\hline
	
	\multirow{4}{*}{\textbf{Allocated \textcolor{red}{$\blacktriangle$}}} & \multirow{4}{*}{\textbf{172}} & \textbf{Non-Null \textcolor{blue}{$\blacktriangle$}}: A null pointer is never valid, not even for accesses of size zero. & \multirow{1}{*}{\textbf{impl}<\textbf{\textcolor{generic}{\textbf{T: Sized}}}> \textbf{NonNull}<\textbf{\textcolor{generic}{\textbf{T}}}>::new\_unchecked}\\
	\cline{3-4}
    & & \textbf{Non-Dangling \textcolor{blue}{$\blacktriangledown$}}: The value must not be pointing to the deallocated memory even for operations of size zero, including data stored in the stack frame and heap chunk. & \multirow{3}{*}{\textbf{trait} \textbf{SliceIndex}<\textbf{\textcolor{generic}{\textbf{T: ?Sized}}}>::get\_unchecked}\\
	\hline
	
	\multirow{6}{*}{\textbf{Bounded \textcolor{red}{$\blacktriangledown$}}} & \multirow{6}{*}{\textbf{141}} & \textbf{Numerical}: The relationship expressions based on numerical operations exhibit clear numerical boundaries. The terms of the expressions can be constants, variables, or the return values of function calls. There are six relational operators including EQ, NE, LT, GT, LE, and GE. & \multirow{4}{*}{\textbf{impl}<\textbf{\textcolor{generic}{\textbf{T: ?Sized}}}> \textcolor{mut}{*mut} \textbf{\textcolor{generic}{\textbf{T}}}::offset\_from}\\
	\cline{3-4}
	& & \textbf{Dereferencable}: The memory range of the given size starting at the pointer must all be within the bounds of a single allocated object. & \multirow{2}{*}{\textbf{impl}<\textbf{\textcolor{generic}{\textbf{T: ?Sized}}}> \textcolor{mut}{*mut} \textbf{\textcolor{generic}{\textbf{T}}}::copy\_from}\\
	\hline
	
	\multirow{7}{*}{\textbf{Initialized \textcolor{red}{$\blacktriangledown$}}} & \multirow{7}{*}{\textbf{104}} & \textbf{Initialized \textcolor{blue}{$\blacktriangle$}}: The value that has been initialized can be divided into two scenarios: fully initialized and partially initialized.  & \multirow{2}{*}{\textbf{impl}<\textbf{\textcolor{generic}{\textbf{T}}}> \textbf{MaybeUninit}<\textbf{\textcolor{generic}{\textbf{T}}}>::assume\_init}\\
	\cline{3-4}
	& & \textbf{Typed \textcolor{blue}{$\blacktriangledown$}}: The bit pattern of the initialized value must be valid at the given type and uphold additional invariants for generics. & \multirow{2}{*}{\textbf{impl}<\textbf{\textcolor{generic}{\textbf{T: ?Sized}}}> \textcolor{mut}{*mut} \textbf{\textcolor{generic}{\textbf{T}}}::read}\\
	\cline{3-4}
	& & \textbf{Encoded \textcolor{blue}{$\blacktriangledown$}}: The encoding format of the string includes UTF-8 string, ASCII string (in bytes), and C-compatible string (nul-terminated trailing with no \texttt{nul} bytes in the middle). & \multirow{3}{*}{\textbf{impl} \textbf{String}::from\_utf8\_unchecked}\\
	\hline
		
	\multirow{7}{*}{\textbf{Layout \textcolor{red}{$\blacktriangledown$}}} & \multirow{7}{*}{\textbf{109}} & \textbf{Sized \textcolor{blue}{$\blacktriangle$}}: The restrictions on Exotically Sized Types (EST), including Dynamically Sized Types (DST) that lack a statically known size, such as trait objects and slices; and Zero Sized Types (ZST) that occupy no space. & \multirow{3}{*}{\textbf{core}::\textbf{mem}::size\_of\_raw}\\
	\cline{3-4}
	& & \textbf{Aligned \textcolor{blue}{$\blacktriangle$}}: The value is properly aligned via a specific allocator or the attribute \texttt{\#[repr]}, including the alignment and the padding. & \multirow{2}{*}{\textbf{impl}<\textbf{\textcolor{generic}{\textbf{T: ?Sized}}}> \textcolor{mut}{*mut} \textbf{\textcolor{generic}{\textbf{T}}}::swap}\\
	\cline{3-4}
	& & \textbf{Fitted \textcolor{blue}{$\blacktriangledown$}}: The layout (including size and alignment) must be the same layout that was used to allocate that block of memory. & \multirow{2}{*}{\textbf{trait} \textbf{GlobalAlloc}::dealloc}\\
	\hline
	
	\multirow{2}{*}{\textbf{SystemIO}} & \multirow{2}{*}{\textbf{26}} & The variable is related to the system IO and depends on the target platform, including TCP sockets, handles, and file descriptors. & \multirow{2}{*}{\textbf{trait} \textbf{FromRawFd}::from\_raw\_fd}\\
	\hline
	
	\multirow{2}{*}{\textbf{Thread}} & \multirow{2}{*}{\textbf{2}} & \textbf{Send}: The type can be transferred across threads. & \multirow{1}{*}{\textbf{core}::\textbf{marker}::\textbf{Send}}\\
	\cline{3-4}
	& & \textbf{Sync}: The type can be safe to share references between threads. & \multirow{1}{*}{\textbf{core}::\textbf{marker}::\textbf{Sync}}\\
	\hline
	
	\multirow{2}{*}{\textbf{Unreachable}} & \multirow{2}{*}{\textbf{5}} & The specific value will trigger unreachable data flow, such as enumeration index (variance), boolean value, etc. & \multirow{2}{*}{\textbf{impl}<\textbf{\textcolor{generic}{\textbf{T}}}> \textbf{Option}<\textbf{\textcolor{generic}{\textbf{T}}}>::unwrap\_unchecked}\\
	\hline
	
	\rowcolor{gray_blank} \multicolumn{4}{|c|}{}\\
	\rowcolor{gray_blank} \multicolumn{4}{|c|}{\multirow{-2}{*}{\textbf{Postcondition Safety Property}}}\\
	\hline
	
	\multirow{6}{*}{\textbf{Aliased}} & \multirow{6}{*}{\textbf{32}} & \textbf{Aliased \textcolor{blue}{$\blacktriangle$}}: The value may have multiple mutable references or simultaneously have mutable and shared references. & \multirow{2}{*}{\textbf{impl}<\textbf{\textcolor{generic}{\textbf{T: ?Sized}}}> \textcolor{mut}{*mut} \textbf{\textcolor{generic}{\textbf{T}}}::as\_mut}\\
	\cline{3-4}
	& & \textbf{Mutated \textcolor{blue}{$\blacktriangledown$}}: The value, which is owned by an immutable binding or pointed by shared reference, may be mutated. & \multirow{2}{*}{\textbf{impl}<\textbf{\textcolor{generic}{\textbf{T: ?Sized}}}> \textcolor{mut}{*const} \textbf{\textcolor{generic}{\textbf{T}}}::as\_ref}\\
	\cline{3-4}
	& & \textbf{Outlived \textcolor{blue}{$\blacktriangledown$}}: The arbitrary lifetime (unbounded) that becomes as big as context demands may outlive the pointed memory. & \multirow{2}{*}{\textbf{impl} \textbf{CStr}::from\_ptr}\\
	\hline
	
	\multirow{2}{*}{\textbf{DualOwned}} & \multirow{2}{*}{\textbf{46}} & It may create multiple overlapped owners in the ownership system that share the same memory via retaking the owner or creating a bitwise copy.  & \multirow{2}{*}{\textbf{impl}<{\textcolor{generic}{\textbf{T: ?Sized}}}> \textbf{Box}<\textcolor{generic}{\textbf{T}}>::from\_raw}\\
	\hline
	
	\multirow{2}{*}{\textbf{Untyped}} & \multirow{2}{*}{\textbf{37}} & The value may not be in the initialized state, or the byte pattern represents an invalid value of its type. & \multirow{2}{*}{\textbf{core}::\textbf{mem}::zeroed}\\
	\hline
	
	\textbf{Freed} & \textbf{19} & The value may be manually freed or automated dropped. & \textbf{impl}<\textbf{\textcolor{generic}{\textbf{T: ?Sized}}}> \textbf{ManuallyDrop}<\textcolor{generic}{\textbf{T}}>::drop\\
	\hline
	
	\textbf{Leaked} & \textbf{35} & The value may be leaked or escaped from the ownership system. & \textbf{impl}<\textbf{\textcolor{generic}{\textbf{T: ?Sized}}}> \textcolor{mut}{*mut} \textcolor{generic}{\textbf{T}}::write\\
	\hline
	
	\textbf{Pinned} & \textbf{5} & The value may be moved, although it ought to be pinned. & \textbf{impl}<\textbf{\textcolor{generic}{\textbf{P: Deref}}}> \textbf{Pin}<\textcolor{generic}{\textbf{P}}>::new\_unchecked\\
    \bottomrule[1pt]
\end{tabular}
\begin{tablenotes}
	\footnotesize
	\item[1] \texttt{Send}~\cite{send} and \texttt{Sync}~\cite{sync} are unsafe traits (markers) that are automatically implemented by the compiler when it determines that they are required. Therefore, they lack associated methods.
	\item[2] The difference between \textit{DualOwned} and \textit{Aliased} is that \textit{DualOwned} only focuses on objects instead of pointers and references.
\end{tablenotes}
\end{threeparttable}
}
\end{table*}

\subsubsection{Pre-condition Safety Property (PRE-SP)} The \textit{PRE-SP} defines an assumption that if the input values do not satisfy any safety requirement, the function call will trigger undefined behavior in the current program point. Thus, any unsafe API can be considered a black box in the single-step execution at the call site, ignoring its implementation~\cite{van1999silicon}. \textit{PRE-SP} aligns with the characteristic of unsafe function, \textit{i.e.}, it cannot ensure safety for arbitrary inputs. In Table~\ref{table:sp}, we define 7 \textit{PRE-SPs} with 14 sub-properties in total. For example, \texttt{ptr::swap}~\cite{swap} has a description in the document slices - \ul{\textit{"Both x and y must be properly aligned."}}, thus categorized into \textit{Layout (Aligned)}. 

\subsubsection{Post-condition Safety Property (POS-SP)} The previous assumption leads to the deduction that the function can be safely called if the proper inputs are supplied. However, this assurance only concerns the safety of the calling site. \textit{POS-SP} focuses on the safety issues that may arise from the subsequent operations, assuming that the input values satisfy all \textit{PRE-SPs}. Table~\ref{table:sp} gives 6 \textit{POS-SPs} with 8 sub-properties. For example, \texttt{mem::zeroed}~\cite{zeored} has a description in the document slices - \ul{\textit{"There is no guarantee that an all-zero byte-pattern represents a valid value of some type T."}}, thus categorized into \textit{Untyped}.

\subsubsection{Isolation} Our assumption established an isolation to improve the soundness of safety properties. It separates the side effects triggered in current call sites or subsequent program points and dissolves the ambiguity - the invalid input will generate the invalid output. Hence, the \textit{POS-SP} can be the superset of \textit{PRE-SP}. Benefiting from this isolation, \textit{PRE-SPs} and \textit{POS-SPs} are not overlapping. It can be verified by a Rust design, where creating raw pointers is always safe, but dereferencing them is unsafe~\cite{reference}. Similarly, we assume that \textit{PRE-SPs} only affect the safety of function calls, while \textit{POS-SPs} focus on the subsequent usage of inputs and return values. Furthermore, \textit{POS-SPs} are only considered under the premise that all \textit{PRE-SPs} are satisfied.

\subsubsection{Corner Cases}
The standard library lists unsafe APIs that cannot be categorized into our current safety properties. We open-sourced the API labels that can be used for indexing~\footnote{https:https://github.com/Artisan-Lab/rustsp\_analyzer} and additional clarifications listed below, which fall into the following categories:

\begin{enumerate}[0]
\item[$\bullet$] \noindent FFI functions with no safety requirements (2), \textit{e.g.,} \texttt{VaListImpl<'f>::arg}~\cite{arg}.
\item[$\bullet$] \noindent Compilation procedure: codegen, intrinsic, and LLVM tag (6), \textit{e.g.,} \texttt{breakpoint}~\cite{breakpoint}.
\item[$\bullet$] \noindent Numerical conversions without an accurate, identifiable boundary (4), \textit{e.g.,} \texttt{nearbyintf32}~\cite{nearbyintf32}.
\item[$\bullet$] \noindent Lacking any explanation of why they are considered unsafe (1), \textit{e.g.,} \texttt{pref\_align\_of}~\cite{prefalignof}.
\end{enumerate}

No API has all safety property labels in the standard library. While it is theoretically feasible to be labeled with all safety properties, the probability is exceedingly low since many safety properties are connected with particular types. For example, \textit{Bounded} is associated with the numerical type. It is incredibly challenging to introduce those complex inputs. However, users can construct an excessively intricate function by hand that requires all safety requirements.

\subsection{Changes in Safety Properties Revision}\label{sec:changes}
We conducted a supplementary revision of the safety properties, namely the fourth-round audit, representing a major improvement. It serves as the foundation for reorganizing the documents. The motivation stems from feedback obtained through user surveys, highlighting that lengthy definitions are detrimental to understanding the safety requirements. The previous version needed to be more competent to introduce subclasses, making it precise for document restructuring. A comparison between the revised version and the earlier version is presented in Table~\ref{table:spc}, and this section will elucidate the reasons for the adjustment.

\begin{table*}[]
\caption{The former version of safety properties that have significant change introduced in the fourth round. The new safety properties have decreased from 19 to 13 major categories, with the addition of secondary subclasses. The safety properties in the gray-shaded text have been restructured in this paper.}
\label{table:spc}
\resizebox{\linewidth}{!}{
\begin{threeparttable}
\begin{tabular}{|c|p{14cm}|c}
    \toprule[1pt]
	\textbf{Safety Property} & \textbf{Definition and the safety requirement of each Safety Property.}\\	
	\hline
	\hline
	\multirow{2}{*}{\textcolor{gray}{\textbf{Const Numeric Bound}}}  & Relational operations allow for compile-time determination of the constant numerical boundaries on one side of an expression, including overflow check, index check, etc.\\
	\hline
	\multirow{2}{*}{\textcolor{gray}{\textbf{Relative Numeric Bound}}} & Relational operations involve expressions where neither side is a constant numeric, including address boundary check, overlap check, size check, variable comparison, etc.\\
	\hline
	\multirow{2}{*}{\textcolor{gray}{\textbf{Dereferencable}}} & The memory range of the given size starting at the pointer must all be within the bounds of a single allocated object.\\	
	\hline
	\hline
   
   	\textcolor{gray}{\multirow{2}{*}{\textbf{Allocated}}} & Value stored in the allocated memory, including data in the valid stack frame and allocated heap chunk, which cannot be \texttt{NULL} or dangling.\\
	\hline
	\hline
    
	\multirow{2}{*}{\textcolor{gray}{\textbf{Initialized}}} & Value that has been initialized can be divided into two scenarios: fully initialized and partially initialized. The initialized value must be valid at the given type (a.k.a. typed).\\
	\hline
	\multirow{2}{*}{\textcolor{gray}{\textbf{Encoding}}} & Encoding format of the string, includes valid UTF-8 string, valid ASCII string (in bytes), and valid C-compatible string (nul-terminated trailing with no nul bytes in the middle).\\
	\hline
	\multirow{4}{*}{\textcolor{gray}{\textbf{Consistent Layout}}} & Restriction on Type Layout, including 1) The pointer's type must be compatible with the pointee's type; 2) The contained value must be compatible with the generic parameter for the smart pointer; and 3) Two types are safely transmutable: The bits of one type can be reinterpreted as another type (bitwise move safely of one type into another).\\
	\hline
	\hline
    
	\multirow{2}{*}{\textcolor{gray}{\textbf{Aligned}}} & Value is properly aligned via a specific allocator or the attribute \texttt{\#[repr]}, including the alignment and the padding of one Rust type.\\
	\hline
	\multirow{2}{*}{\textcolor{gray}{\textbf{Exotically Sized Type}}} & Restrictions on Exotically Sized Types (EST), including Dynamically Sized Types (DST) that lack a statically known size, such as trait objects and slices; Zero Sized Types (ZST) that occupy no space.\\
	\hline
	\hline
   
	\multirow{3}{*}{\textcolor{gray}{\textbf{Aliasing \& Mutating}}} & Aliasing and mutating rules may be violated, including 1) The presence of multiple mutable references; 2) The simultaneous presence of mutable and shared references, and the memory the pointer points to cannot get mutated (frozen); 3) Mutating immutable data owned by an immutable binding.\\
	\hline
	\multirow{2}{*}{\textcolor{gray}{\textbf{Outliving}}} & Arbitrary lifetime (unbounded) that becomes as big as context demands or spawned thread, may outlive the pointed memory.\\		
    \bottomrule[1pt]
\end{tabular}
\end{threeparttable}
}
\end{table*}

\subsubsection{\textbf{\textit{New: Bounded.}}}\textit{Bounded} has two subclasses. The separation between \textit{Numerical} and \textit{Dereferencable} achieves better independence for numerical calculations and memory access. \textit{Numerical} arises from \textcolor{gray}{\textit{Const Numeric Bound}} and \textcolor{gray}{\textit{Relative Numeric Bound}}. In user surveys, participants pointed out that this separation lacked significant differences because rearranging inequalities can result in constant and non-constant boundaries. The merged definition focuses on relational expressions with numerical boundaries, as most numerical conditions can be checked using \texttt{assert!()}. It also expands the scope of terms, supporting constants, variables, and function calls in expressions. \textit{Dereferencable} is the second subclass. In the previous version, the scope among \textcolor{gray}{\textit{index check}}, \textcolor{gray}{\textit{address boundary check}}, and \textcolor{gray}{\textit{overlap check}} were easily misunderstanding. They all regulate the valid offset for the memory access but can be covered by \textit{Dereferencable}.

\subsubsection{\textbf{\textit{New: Allocated.}} } The official document gives the description of \textit{valid pointers} (not strictly defined), which is the basis of \textit{Allocated} in both versions. The changes made to \textit{Allocated} are minor, primarily because \textit{Non-Null} and \textit{Non-Dangling} have a hierarchical relationship, where satisfying \textit{Non-Null} is a prerequisite for satisfying \textit{Non-Dangling}. Many APIs accept pointers to be \textit{NULL} or dangling because it will be overwritten. For others, if they accept a non-null pointer, users must ensure that it cannot be dangling, as the pointer may be used to access memory, which would result in a memory safety issue.

\subsubsection{\textbf{\textit{New: Initialized.}}}\textit{Initialized} restructures from \textcolor{gray}{\textit{Initialized}}, \textcolor{gray}{\textit{Consistent Layout}}, and \textcolor{gray}{\textit{Encoding}}. It results in three new subclasses: \textit{Initialized}, \textit{Typed}, and \textit{Encoded}. They form a hierarchical relationship, namely ensuring \textit{Typed} must guarantee \textit{Initialized}, and ensuring \textit{Encoded} must guarantee \textit{Typed}. Since the initialization status is a fundamental prerequisite, the initialized value may not meet all safety requirements for a specific type. It assesses whether the initialized bits can be translated into a valid value for the given type. \textcolor{gray}{\textit{Consistent Layout}} highlights the type of the pointed-to region, the generic parameters of containers, and the safe conversion of types—all of which are essentially \textit{Typed} problems. \textit{Encoded} introduces a third level for string format, building on the prerequisites set by \textit{Typed}. We give a prominent example of converting \texttt{Vec<u8>} to \texttt{String}. It encompasses multiple \textit{Initialized} requirements, including initialized state, typed state, and string format.

\subsubsection{\textbf{\textit{New: Layout.}}}\textit{Layout}, being a consolidation of \textcolor{gray}{\textit{Aligned}}, and \textcolor{gray}{\textit{Exotically Sized Types (EST)}}, is unrelated to \textcolor{gray}{\textit{Consistent Layout}}. \textcolor{gray}{\textit{Consistent Layout}} emphasizes type, representing requirements for initialized status and the ability to safely cast. \textit{Layout} restricts adaptability to memory blocks, specifying size and alignment for underlying data. Note that it is independent of types; it solely considers allocated blocks' size and alignment without considering whether the internal bits should meet any type-specific requirements. This difference is also emphasized in contrast to \textit{Initialized}. Among the subclasses, \textit{Fitted} needs to satisfy both size and alignment requirements, and \textit{Sized} focuses on restrictions for the ESTs.

\subsubsection{\textbf{\textit{New: Aliased.}}} The changes to \textit{Aliased} resulted from merging \textcolor{gray}{\textit{Aliasing \& Mutating}} and \textcolor{gray}{\textit{Outliving}}. In the previous version of API labels, we observed a high correlation between these safety properties, often appearing together in labels. Similar to \textcolor{gray}{\textit{Consistent Layout}}, feedback from survey participants on \textcolor{gray}{\textit{Aliasing \& Mutating}} emphasized the description being overly lengthy, prompting us to break it down. In the ownership model, \textit{Aliased}, \textit{Mutated}, and \textit{Outlived} form a triple restriction for memory-safe access. \textit{Aliased} pertains to restrictions on the exclusiveness of access permissions, \textit{Mutated} involves limitations on access operations, and \textit{Outlived} imposes constraints on the legitimate program points for access actions. The subdivided classes align more closely with the safety constraints of the ownership system.

\subsection{Correlation Analysis on Labeled API} Although one of our design goals focuses on non-overlapping, it is necessary to investigate potential correlations inside those safety properties. This notion is from the intuition that data satisfying \textit{Dereferenceable} in \textit{Bounded} should always meet \textit{Allocated}.

\begin{figure*}[]
\begin{subfigure}[]{0.49\textwidth}
	\includegraphics[width=\textwidth]{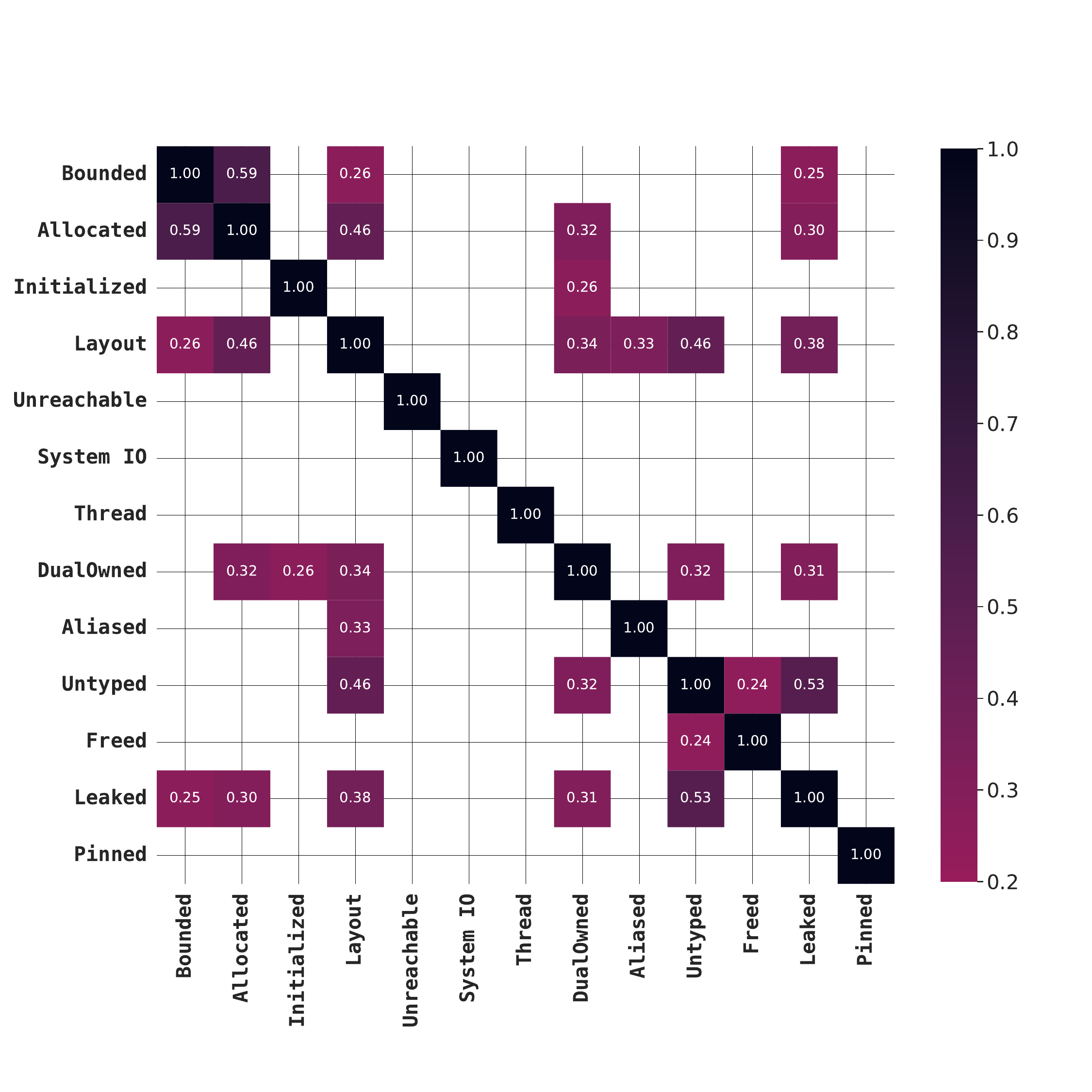}
	\caption{Analysis results on the large dataset (original).}
	\label{fig:co1}
\end{subfigure}
\begin{subfigure}[]{0.49\textwidth}
	\includegraphics[width=\textwidth]{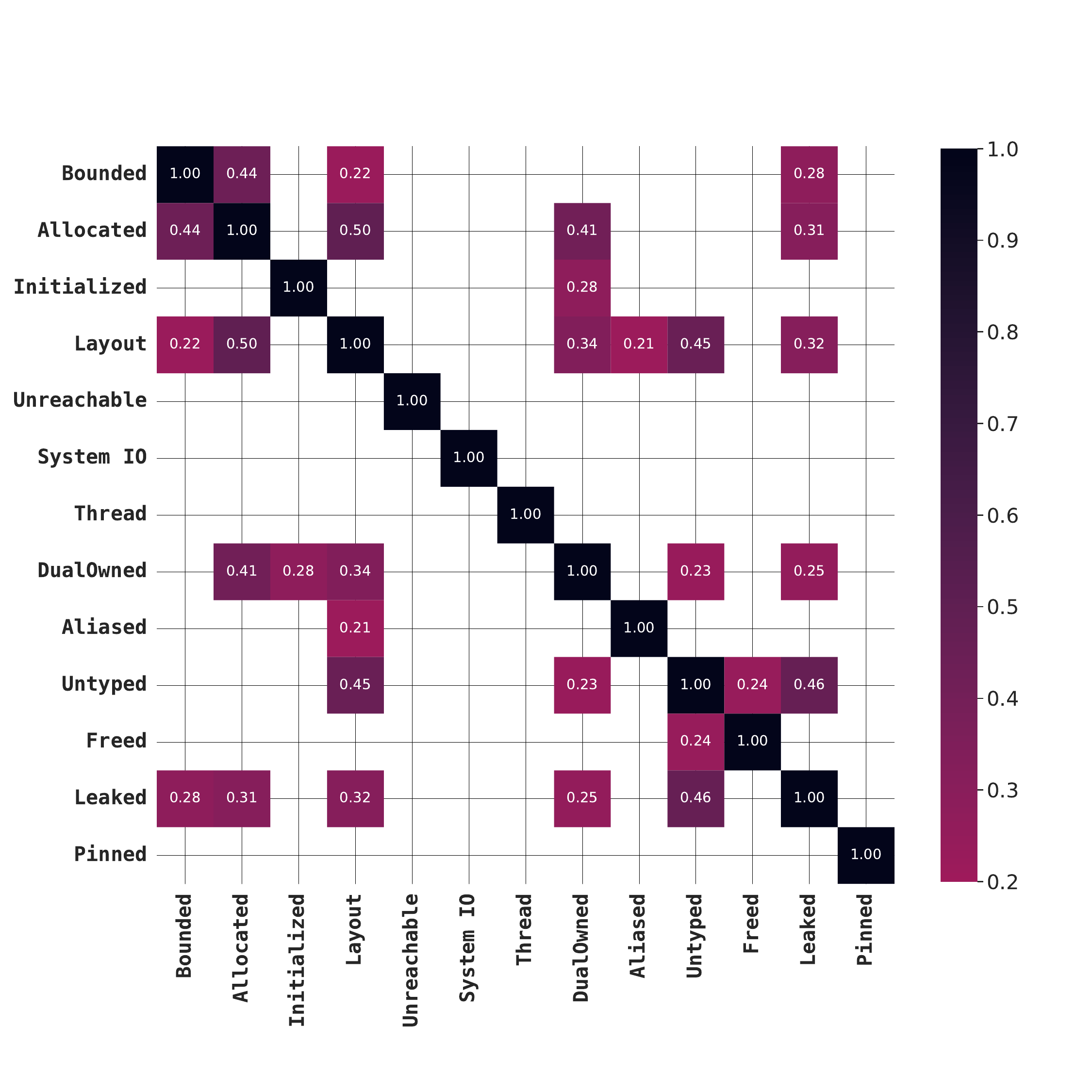}
	\caption{Analysis results on the small dataset (filtered).}
	\label{fig:co2}
\end{subfigure}
\caption{Correlation matrices for both the large and small datasets. Each figure only includes the sections with weak correlation and above (correlation greater than 0.20).}
\label{fig:co}
\end{figure*}

\subsubsection{Methodology} We created two databases based on the labeled APIs in the different scales and then conducted a correlation analysis. The analysis results demonstrate the anticipated differences.

\noindent \underline{\textit{Large Dataset.}} The large dataset adopts the original collection, including the entire set of unsafe APIs with labels. The intent of keeping a large dataset is to emphasize the quantity because functionally related APIs may have similar labels, and adequate data can expose potential correlations.

\begin{filter}\label{filter:small}
APIs must have the same labels and satisfy at least one of the following requirements:
\begin{enumerate}[0]
\item[$\bullet$] \noindent Implementations of the same method with different mutability.
\item[$\bullet$] \noindent Implementations of the same trait for different types, including mono-morphizations in the trait or struct definitions.
\item[$\bullet$] \noindent Functions with the same name implemented for different types within the same namespace.
\item[$\bullet$] \noindent Encapsulation of intrinsic functions.
\end{enumerate}
\end{filter}

\noindent \underline{\textit{Small Dataset.}} The small dataset is created by applying Filter~\ref{filter:small} to the large dataset. This is done to counteract the potential bias from excessive similar APIs. The small dataset intends to eliminate the redundancy of irrelevant data and concentrate on diversity.

\subsubsection{Correlation Matrix} In Figure~\ref{fig:co}, we built correlation matrices for two datasets, retaining the elements with moderate correlation and above (correlation coefficient > 0.20). The large dataset has a higher susceptibility to interference from redundant APIs. For example, there are 30 implementations of the trait \texttt{SliceIndex<[T]>}~\cite{sliceindex}, all of which are labeled with \textit{Bounded} and \textit{Allocated} only. The correlation between them is higher in the large dataset, changing from 0.44 to 0.59. In the small dataset, the diverse functionality is more likely to result in the loss of pertinent data that could affect correlations. For example, the small dataset's correlation between \textit{Allocated} and \texttt{DualOwned} changed from 0.32 to 0.41. At last, \textit{Unreachable}, \textit{System IO}, \textit{Thread} and \textit{Pinned} achieve the best independence, as they have no substantial correlation with any other safety properties in both matrices.

\textbf{\textit{New Findings in Correlation Matrix.}} Based on two diagrams in Figure~\ref{fig:co}, we extracted all the pairs with at least a moderate CC, as listed in Table~\ref{table:pair}. Here, we will provide a more comprehensive explanation of the improved correlations. It is evident that the new safety property classification exhibits better independence. Notably, the pairs have significantly decreased, going from 11 to 4 pairs. The main reason is the reduction in major categories from 19 to 13 and the addition of secondary subclasses.

The case studies learned in the previous work are interpretable~\cite{cui2024unsafe}, which summarized six requirements for dereferencing: \textcolor{gray}{\textit{Allocated}}, \textcolor{gray}{\textit{Relative Numeric Bound}}, \textcolor{gray}{\textit{Dereferencable}}, \textcolor{gray}{\textit{Consistent Layout}}, \textcolor{gray}{\textit{Aligned}}, and \textcolor{gray}{\textit{Initialized}}. It can be inferred that the correlation between \textit{Allocated} and \textit{Bounded} is due to the merging \textcolor{gray}{\textit{Relative Numeric Bound}} and \textcolor{gray}{\textit{Dereferencable}} into \textit{Bounded}. Because we remove memory range issues from \textcolor{gray}{\textit{Relative Numeric Bound}}, \textit{Allocated} is highly correlated with the subclass \textit{Dereferencable} in \textit{Bounded}. The correlation between \textit{Allocated} and \textit{Layout} is the result of merging \textcolor{gray}{\textit{Aligned}} into \textit{Layout}. Therefore, transitioning to the new safety properties, the safety requirements of dereferencing can be updated to include: \textit{Allocated}, \textit{Bounded}, \textit{Layout}, and \textit{Initialized}.

\subsubsection{Hierarchical Relation} Hierarchical relationships were observed for both primary and secondary classes, as annotated by the dependency conditions in Table~\ref{table:sp}. It stems from expert experience and analysis results of the correlation matrix. At the primary level, \textit{Allocated} is a first-class member, which is necessary for three safety properties. Similarly, there exist hierarchical relationships among sub-classes. For a pointer to be \textit{Non-Dangling}, it must first satisfy the condition of being \textit{Non-Null}. Notably, the hierarchical relationships among subclasses do not propagate to the parent classes, whereas the parent classes can propagate to subclasses.

\begin{table}[]
\caption{SP pairs with correlation coefficients (CC) greater than 0.4 in the large and small dataset correlation matrices.}
\label{table:pair}
\resizebox{0.55\linewidth}{!}{
\begin{tabular}{|cc|>{\columncolor{gray_item}}c|cc|>{\columncolor{gray_item}}c|}
    \toprule[1pt]
	\textbf{SP1} & \textbf{SP2} & \textbf{AVG-CC} & \textbf{SP1} & \textbf{SP2} & \textbf{AVG-CC} \\
	\hline
	Allocated       & Bounded          & 0.51  & Allocated       & Layout           & 0.48  \\
	\hline
	Layout          & Untyped          & 0.45  & Leaked          & Untyped          & 0.49  \\
    \bottomrule[1pt]
\end{tabular}
}
\end{table}

\subsection{Revising Safety Property Documents} After obtaining the refined safety property definitions and corresponding API labels, this section aims to reorganize the Unsafe Rust documents based on the existing safety properties, providing a developer-friendly form. An API was only marked as requiring or not requiring fulfillment of specific safety properties in the previous work, disregarding further details~\cite{cui2024unsafe}. In this paper, we aim to break through this barrier. The process has the following goals:

\begin{goal}
	\textbf{Independence}: Each unsafe API has a unique identifier maintained by one Safety Property document.
\end{goal}

\begin{goal}\label{g:sp1}
	\textbf{Intuitiveness}: Safety Property documents cannot contain hyperlinks referencing other documents' safety requirements.
\end{goal}

\begin{goal}\label{g:sp2}
	\textbf{Comprehensiveness}: Safety requirements may be distributed across different locations; efforts should be made to cover the entire native document and delete the irrelevant content.
\end{goal}

\begin{goal}\label{g:sp3}
	\textbf{Completeness}: Incomplete safety descriptions within the native document should be supplemented as much as possible.
\end{goal}

\begin{goal}
	\textbf{Structurality}: Safety requirements are enumerated in a structured manner and presented to users in a clear and logically organized format.
\end{goal}

\begin{figure}[]
\begin{subfigure}[]{\textwidth}
\begin{lstlisting}[style=docs, label=list:read_native, caption=Native document of ptr::read in Rust 1.75.\\] 
# Functionality Description
Reads the value from src without moving it. This leaves the memory in src unchanged.

# Safety Description
Behavior is undefined if any of the following conditions are violated:
src must be valid for reads.
src must be properly aligned. Use read_unaligned if this is not the case.
src must point to a properly initialized value of type T.
Note that even if T has size 0, the pointer must be non-null and properly aligned.

# Subchapter: Ownership of the Returned Value
read creates a bitwise copy of T, regardless of whether T is Copy. If T is not Copy, using both the returned value and the value at *src can violate memory safety. Note that assigning to *src counts as a use because it will attempt to drop the value at *src.
write() can be used to overwrite data without causing it to be dropped.
}
\end{lstlisting}
\end{subfigure}

\begin{subfigure}[]{\textwidth}
\begin{lstlisting}[style=docs, label=list:read_sp, caption=Safety property document of ptr::read in Rust 1.75.\\]
# SafetyProperty
# Identifier: primitive.pointer.html#method.read
# Type: *const T
# Signature: fn read<T>(src: *const T) -> T

self: Allocated
A null pointer is never valid, not even for accesses of size zero.
Even for operations of size zero, the pointer must not be pointing to deallocated memory.

self: Bounded
The memory range of the given size starting at the pointer must all be within the bounds of a single allocated object.

self: Initialized
self must point to a properly initialized value of type T.

self: Layout
self must be properly aligned.

retval: DualOwned
If T is not Copy, using both the returned value and the value at self can violate memory safety. Note that assigning to self counts as a use because it will attempt to drop the value at self.
\end{lstlisting}
\end{subfigure}
\caption{The example of a revised document compared with the native document for unsafe API \texttt{ptr::read}. The safety property document has a unique identifier and matches the safety properties, parameters, and native document slices, providing a more structured format. Additionally, it eliminates the secondary link jumps, making it more intuitive.}
\label{fig:read}
\end{figure}

We established a revision process driven by the above goals. We extract each API's signature, safety property labels, and native documents. Native documents encompass functional descriptions, safety descriptions, and all subchapters. The revised document uses the online \texttt{Rustdoc} link as the unique identifier, allowing each API to redirect to a distinct webpage. Given challenges such as identical function names, methods in traits, and implementations for different mono-morphizations, our identifiers provide better recognizability.

For the native documents, we split the text into sentences and determine whether each sentence can be included in the safety properties. If it can be included, we identify the related parameter (return value), forming a triplet (\textit{i.e.,} parameter $P$, safety property $S$, document slice $D$). To achieve Goal~\ref{g:sp1} to \ref{g:sp3}, we flatten the linked documents and manually supplement missing descriptions, drawing from other slices in the standard library. Figure~\ref{fig:read} illustrates the revision for \texttt{ptr::read}~\cite{read}. The safety property document, compared to the native document, exhibits a more structured format.

\subsection{Integrating Revised Documents into Rust Analyzer}
 rust-analyzer is a widely used Language Server Protocol (LSP) for the Rust programming language. It communicates with VSCode through a background process to provide features such as code completion and navigation. In real-world Rust programming, we have integrated safety property documents into rust-analyzer to assist users in reading and using unsafe functions more safely. This section introduces the overall framework of our prototype.
 
\begin{figure}[]
\includegraphics[width=\textwidth]{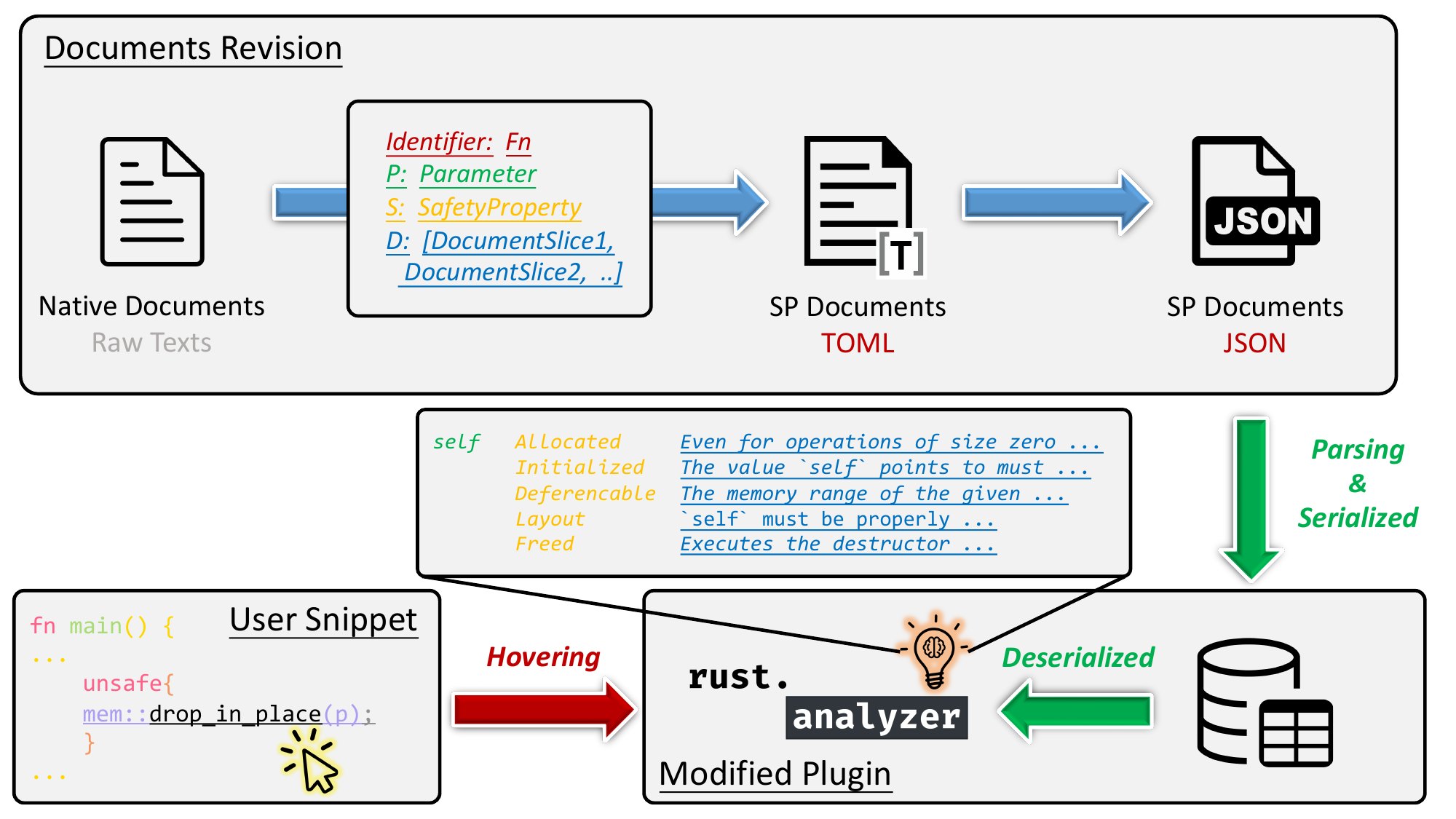}
\caption{The workflow of the plugin. This tool implants safety property documents into the hover component of the rust-analyzer, providing users with safety prompts when hovering over unsafe functions.}
\label{fig:plugin}
\end{figure}

In Figure~\ref{fig:plugin}, the safety properties documents are encoded in a TOML file, using key-value pairs to store the entire set. Each unsafe API has a unique identifier as the primary key, forming the basis for document maintenance. The TOML file adopts the following recording method: 

\begin{center}
	 \texttt{Parameter.SafetyProperty.[ DocumentSlice1, DocumentSlice2, .. ]}
\end{center}

 \noindent Since one safety property may connect with several parameters, one parameter may have to satisfy multiple safety properties, and one safety property for one parameter may correspond to various document slices; the primary key is the parameter, the secondary key is safety property, and the tertiary key is an array of document slices in the markdown format.
 
We convert TOML into JSON, use an external program to parse it and serialize it in a specific data structure. It records the mapping between API identifiers and safety property documents. The analyzer can deserialize the mapping to avoid parsing overhead. In rust-analyzer, we customized the hover component; the hover component can show additional information, such as the document for a definition, when "focusing" code. Focusing is usually done by hovering with a mouse but can also be triggered with a shortcut. When a user hovers over a function, we generate a link to the webpage document for that function to match our identifier. If the database contains its safety property document, we intercept the hover component, inject the safety property document, and callback to the user.

\section{Verifying Real-world Program Vulnerabilities}\label{sec:cveio}
To answer RQ3, this section presents the practical usability of safety properties in real-world scenarios by analyzing the root cause of existing CVEs to validate SP coverage. It also surveys the frequency of unsafe API in the Rust ecosystem by gathering the unsafe API usage on crates.io~\cite{cratesio}.

\begin{figure}[]
\begin{subfigure}[]{\textwidth}
\begin{lstlisting}[language=Rust, style=colouredRust, label=list:cvecode, caption=Source code of CVE-2021-45709 in the crypto2 crate through 2021-10-08 for Rust.\\]
#[inline]
fn xor_si512_inplace(a: &mut [u8], b: &[u32; Chacha20::STATE_LEN]) {
    unsafe {
        let d1 = core::slice::from_raw_parts_mut(a.as_mut_ptr() as *mut u32, Chacha20::STATE_LEN);
        for i in 0..Chacha20::STATE_LEN {
            d1[i] ^= b[i];
        }
    }
}
\end{lstlisting}
\label{list:cvecode}
\end{subfigure}

\begin{subfigure}[]{\textwidth}
\begin{lstlisting}[style=docs, label=list:cvedoc, caption=Description documented in RUSTSEC-2021-0121.\\]
Description
The implementation does not enforce alignment requirements on input slices while incorrectly assuming 4-byte alignment through an unsafe call to std::slice::from_raw_parts_mut, which breaks the contract and introduces undefined behavior.
\end{lstlisting}
\label{list:cvedesp}
\end{subfigure}
\caption{Example of the CVE classification. The buggy source code and description of CVE-2021-45709 (in RUSTSEC). This CVE violates the safety requirement of \textit{Layout: Aligned} and triggers UB when using the unsafe API slice::from\_raw\_parts\_mut.}
\label{fig:cveexample}
\end{figure}

\subsection{Studying Root Causes of Rust CVEs}\label{sec:cve}

\subsubsection{Workflow} We first create a database of existing CVEs filtered by misusing unsafe code and then classify their root causes into safety properties by manual code review.

\noindent \underline{\textit{CVE Database.}} We utilized the CVEMitre~\cite{cve} as the dataset. To retain Rust-related CVEs as many as possible, we conducted searches by keyword \textit{\textbf{"Rust"}}. The results were sorted in reverse chronological order of CVE IDs, ranging from CVE-2017-20004~\cite{cve20004} to CVE-2024-21629~\cite{cve21629}. We filtered them through the CVE descriptions to only include those involving memory safety issues, removing others such as information leaks. Additionally, we filtered CVEs with panic safety since the security implications during stack unwinding are highly dependent on where the panic occurs~\cite{rustonomicon}. Furthermore, retained CVEs should not be associated with deprecated or yanked crates and should be linkable to source code for code audits.

\begin{figure}[]
\includegraphics[width=0.7\textwidth]{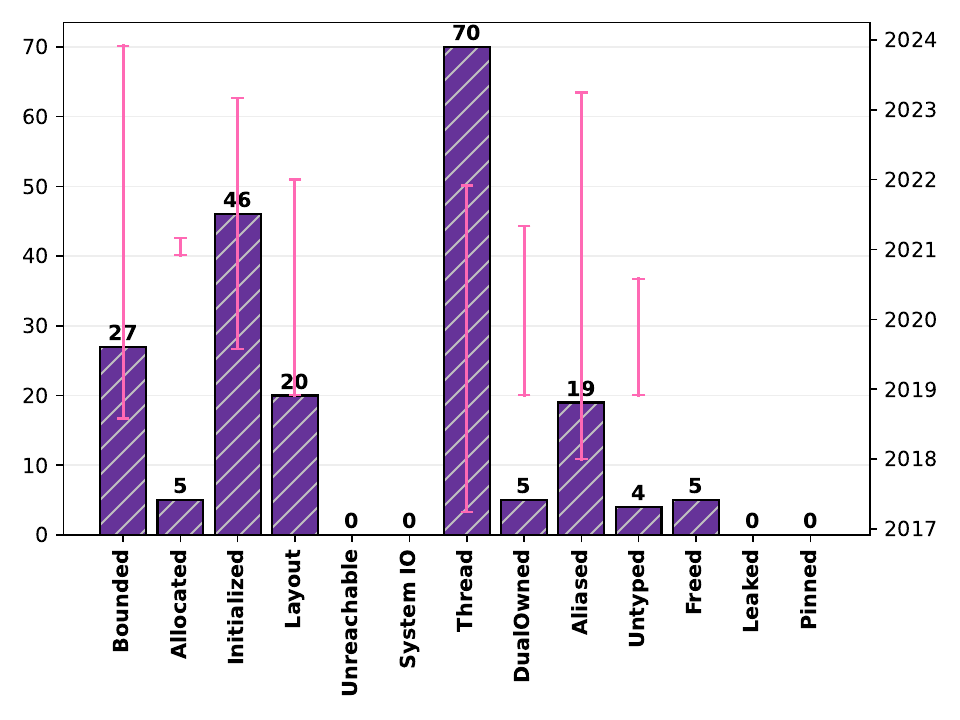}
\caption{SP classification results and their duration on existing Rust CVEs. These CVEs are memory-safety issues resulting from misusing unsafe code and ignoring unrelated issues, such as leaking sensitive information.}
\label{fig:cve}
\end{figure}

\noindent \underline{\textit{CVE Audit.}} We conducted an in-depth analysis of the retained CVEs, utilizing manual code auditing to determine whether the erroneous code segments were caused by misuse of unsafe code, further analyzing whether they could be classified as violations of safety properties. The first and second authors double-checked the results. Due to the concise descriptions provided on the CVEMitre, we considered multiple sources, including issues, pull requests, and RUSTSEC~\cite{rustsec}. All CVEs caused by safe code were excluded, but not those originating from unsafe interiors. In the final dataset (198 CVEs in total), 86.36\% of them were attributed to misuse of unsafe APIs within the standard library. The remaining CVEs were caused by other unsafe operations (\textit{e.g.,} dereferencing raw pointers), usage of unsafe functions outside the standard library, and issues occurring beyond Rust's FFI boundaries.

\noindent \underline{\textit{CVE Example.}}We present our analysis of a typical CVE. Figure~\ref{fig:cveexample} provides the necessary details for auditing CVE-2021-45709~\cite{cve45709,sec0121}. Based on the explicit error description, we were able to locate the defective source code and confirm the misuse of \texttt{slice::from\_raw\_parts\_mut}~\cite{fromptrmut}. Specifically, it was found that the mismatch between the constraints of the parameter \texttt{a} (line 2) and the unsafe call (line 4) resulted in undefined behavior. In our labeled results, this API is marked with the following safety properties: \textit{Allocated}, \textit{Bounded}, \textit{Initialized}, \textit{Layout}, and \textit{Aliased}. This CVE violated the requirement of \textit{Layout: Aligned}, leading to undefined behavior.

\subsubsection{Results} We screened 419 CVEs and conducted thorough code reviews on the remaining 198 CVEs after filtering. We categorized them based on the root causes, matching them with safety properties, and analyzed their distribution. Manual validation confirmed that the root causes of CVEs caused by unsafe Rust did not exceed the scope of safety properties we have defined.

\noindent \underline{\textit{Distribution.}} Figure~\ref{fig:cve} depicts the results: \textit{Thread} (70), \textit{Initialized} (46), and \textit{Bounded} (27) has a significant number. Following are \textit{Layout} (20), \textit{Aliased} (19), \textit{Allocated} (5), and the other three safety properties (ranging from 4 to 20). 4 safety properties have no corresponding CVEs. The period of most CVEs is gathered from as early as August 2019 to as late as December 2021 in a temporal perspective, which contains 91.84\% of listed CVEs.

\noindent \underline{\textit{Case Study.}} The categorized results reveal that the highest proportion is caused by violations of \textit{Thread} (70). It caused by the incorrect \texttt{Send}/\texttt{Sync}~\cite{send,sync} implementations for the user-defined types, without verifying whether each field meets the requirements of those traits, potentially leading to data races and memory safety issues across thread boundaries. \textit{Initialized} ranked second (46), with scenarios summarized as follows: 1) creating uninitialized buffers and passing them as parameters to custom \texttt{Read}~\cite{read} implementations, exposing uninitialized memory to safe code; 2) modifying the field directly to increase buffer lengths in vectors, leading to out-of-bounds writes and dropping uninitialized memory; 3) creating an uninitialized \texttt{NonNull} pointer, violating data invariance.

\subsection{Gathering Statistics on crates.io Ecosystem}\label{sec:cratesio}
Section~\ref{sec:cve} indicates that 86.36\% of the retained CVEs bounded with safety properties were caused by misusing unsafe APIs from the standard library. This observation prompted us to conduct a statistical analysis of unsafe API usage within the Rust ecosystem.

\begin{figure}[]
\includegraphics[width=0.7\textwidth]{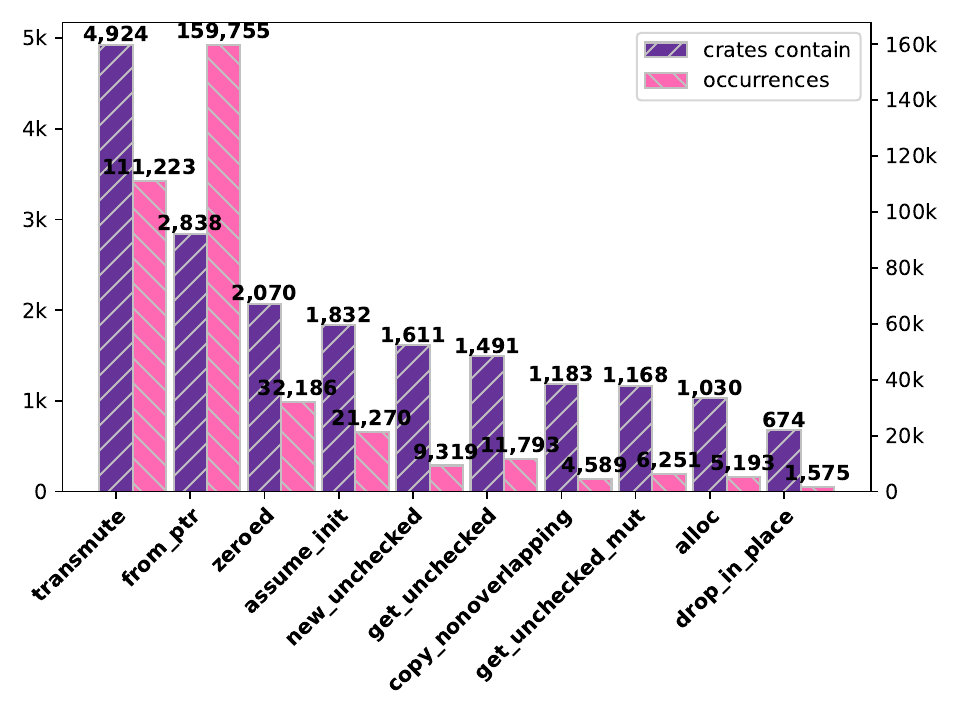}
\caption{Statistics on unfiltered strings (unsafe APIs) in the crates.io ecosystem. The top ten most frequently used strings across all repositories and their source code occurrences are sorted by the sum of crates.}
\label{fig:crates.io}
\end{figure}

\subsubsection{Crates.io Database} Crates.io is the package platform for Rust, where open-source libraries can be published for users. We utilized all active repositories on this platform as our database and employed regular expression matching on the source code to gather statistics, a process that does not involve compilation to aviod overhead. Based on function names, we merged identically named unsafe APIs, creating a dictionary containing 140 unique strings. We removed 20 strings that were identical to other safe functions in the standard library (\textit{e.g.}, \texttt{add}~\cite{addsafe,addunsafe}). To enhance the accuracy, we further filtered out cases where the \textit{unsafe} keyword was not used in the source code.

\subsubsection{Statistics} As of 2023-01-30, we mined all latest crates. The statistics indicate that 21,506 crates contain unsafe code among the 103,516 crates (3,614 are yanked). We generated a summary for each string: the number of crates for the string appearance and the total usage across all crates. The top ten most frequently used strings are shown in Figure~\ref{fig:crates.io}, which depicts statistical results in two dimensions. We observed that the primary scenarios encompass type conversions (\texttt{transmute}), manual memory management (\texttt{zeroed}, \texttt{alloc}, \texttt{drop\_in\_place}), unsafe constructors (\texttt{new\_unchecked}), deferred initialization (\texttt{assume\_init}), unsafe indexing (\texttt{get\_unchecked/mut}), unsafe referencing (\texttt{from\_ptr}), and unsafe memory copying (\texttt{copy\_nonoverlapping}).

\section{Threats to Validity}
For internal validity threats, the standard library might not provide exhaustive coverage for investigating the categories toward safety requirements. We validated the results based on the current CVE to address this limitation. Based on expert knowledge, different programmers may hold divergent views regarding which sub-properties of safety properties should be merged or separated; thus, no fixed conclusion can be drawn. The primary contribution is to provide a method to extract safety properties, label functions, and revise documents. There may be other methods besides ours, but our research is the inaugural study to concentrate on categorizing safety requirements in Rust.

For external validity threats, due to the ongoing development, the programming style and the language standard may evolve. Future updates may introduce new unsafe APIs, deprecate old APIs, or modify them. We acknowledge the existence of uncommon safety descriptions that cannot be classified based solely on the standard library or CVEs because they have not been documented yet.

\section{Related Work}
\textbf{\textit{Empirical Studies on Rust.}} Empirical studies focus on how developers write unsafe code in real-world programs~\cite{astrauskas2020programmers, evans2020rust, qin2020understanding, unsafesyntactic, yu2019fearless} or existing CVEs~\cite{xu2021memory}. They summarize bug patterns and insights into different aspects of safety guarantees. However, these studies do not classify the safety descriptions through the documents. Several empirical studies focus on the Rust learning curve~\cite{abtahi2020learning, fulton2021benefits} and the programming challenges introduced by compiler errors~\cite{zhu2022learning}. Researchers also leveraged the Stack Overflow comment to understand real-world development problems~\cite{zhu2022learning}. However, we are more concerned with experienced system engineers who are proficient than Rust beginners. They need Unsafe Rust to achieve low-level control and better understand the safety requirements when writing unsafe code.

\noindent \textbf{\textit{Bug Detections in Rust.}}
Researches use different methods to find bugs in Rust programs, including formal verification~\cite{lattuada2023verus, ho2022aeneas, astrauskas2022prusti, dang2019rustbelt, jung2019stacked, jung2017rustbelt, wolff2021modular, matsushita2021rusthorn, hahn2016rust2viper, crichton2022modular}, symbolic execution~\cite{lindner2018no, mirai}, model checking~\cite{toman2015crust, vanhattum2022verifying, bae2021rudra, li2021mirchecker, cui2023safedrop}, interpreter~\cite{miri}, and fuzzing~\cite{dewey2015fuzzing, jiang2021rulf}. Some of the above approaches provide bug patterns corresponding to safety properties, as listed in Table~\ref{table:tool}. For example, a significant portion of CVEs on \textit{Initialized} (85.3\%) and \textit{Thread} (92.9\%) were found by \textsc{Rudra}~\cite{bae2021rudra}. This observation suggests that analyzers designed for bug patterns may effectively identify vulnerabilities that violate specific safety properties. Many bugs related to safety properties may have yet to be discovered, as shown in Figure~\ref{fig:cve}.

\begin{table}[]
\caption{Open-source code analyzers that can detect specific safety properties. Each static bug detection tool may not necessarily support all scenarios related to the corresponding safety property.}
\label{table:tool}
\resizebox{0.7\linewidth}{!}{
\begin{tabular}{cp{8cm}}
    \toprule[1pt]
	\textbf{Static Analyzer} & \textbf{Supported Safety Properties}\\
	\midrule[1pt]
	\textbf{\textsc{Rudra}}~\cite{bae2021rudra} & \textit{Initialized}, \textit{Thread}, \textit{DualOwned}\\
	\midrule[1pt]
	\textbf{\textsc{SafeDrop}}~\cite{cui2023safedrop} & \textit{Allocated}, \textit{Initialized},\textit{DualOwned}, \textit{Freed}\\
	\midrule[1pt]
	\textbf{\textsc{MirChecker}}~\cite{li2021mirchecker} & \textit{Bounded}, \textit{DualOwned}\\
	\midrule[1pt]
	\textbf{\textsc{FFIChecker}}~\cite{li2022detecting} & \textit{Allocated}, \textit{Leaked} \textit{(Rust/C FFI only, based on LLVM)} \\
	\midrule[1pt]
	\textbf{\textsc{rCanary}}~\cite{rCanary} & \textit{Leaked} \\
    \bottomrule[1pt]
\end{tabular}
}
\end{table}  

\noindent \textbf{\textit{Benefiting Rust Community.}}
CVE classification provides a set of CVE lists for each safety property that can be used as a benchmark. This benchmark can be employed to evaluate the effectiveness of research prototypes or bug-detection tools designed for particular safety properties (\textit{e.g.,} \texttt{SafeDrop} and \texttt{Rudra}). As far as we know, the Rust community needs a unified, ground-truth-supported benchmark to support effectiveness comparisons based on safety issues. We advocate for setting such a benchmark to serve as a basis for the SE community.

\section{Conclusion}
With increasing system programs adopting Rust, understanding the safety requirements when writing unsafe code is crucial, particularly with well-defined categorization. To this end, we conducted a comprehensive empirical study on safety requirements through the unsafe boundary. In this paper, we extended and refined our study in ICSE 2024. We focus on unsafe API documents in the standard library to infer safety properties and then categorize unsafe APIs and existing CVEs. Furthermore, we completely revised the native documents and integrated them into the rust-analyzer. This tool prompts users with safety requirements when invoking unsafe APIs. Through these efforts, we aim to promote the standardization of systematic documents for Unsafe Rust in the Rust community.

\section*{ACKNOWLEDGMENTS}
This work is supported by the National Natural Science Foundation of China (No. 62372304).

\bibliographystyle{ACM-Reference-Format}
\bibliography{sp}

\end{document}